\documentclass[11pt]{article}
\usepackage{amsmath,amssymb}
\usepackage{array}
\newcount\figureno     \figureno=0
\newdimen\figdim       \figdim=70mm
\def\figureinc{%
   \global\advance\figureno by 1%
}
\def\figcaption#1#2#3{\hbox to #2{\hss{\vbox{\hsize=#2 \parindent=0pt
        {\bf Figure \number\figureno#3 :\ }#1}}\hss}
}

\begin{document}
\baselineskip 100pt
\renewcommand{\baselinestretch}{1.5}
\renewcommand{\arraystretch}{0.666666666}
{\large
\parskip.2in
\numberwithin{equation}{section}
\newcommand{\be}{\begin{equation}}
\newcommand{\ee}{\end{equation}}
\newcommand{\ben}{\begin{equation*}}
\newcommand{\een}{\end{equation*}}
\newcommand{\eqalinb}{\begin{eqnarray}}
\newcommand{\eqaline}{\end{eqnarray}}
\newcommand{\br}{\bar}
\newcommand{\fr}{\frac}
\newcommand{\lm}{\lambda}
\newcommand{\ra}{\rightarrow}
\newcommand{\al}{\alpha}
\newcommand{\bt}{\beta}
\newcommand{\z}{\zeta}
\newcommand{\pa}{\partial}
\newcommand{\hs}{\hspace{5mm}}
\newcommand{\up}{\upsilon}
\newcommand{\bigb}{\hspace{7mm}}
\newcommand{\dg}{\dagger}
\newcommand{\vphi}{\vec{\varphi}}
\newcommand{\ve}{\varepsilon}
\newcommand{\acc}{\\[3mm]}
\newcommand{\dl}{\delta}
\newcommand{\sdil}{\ensuremath{\rlap{\raisebox{.15ex}{$\mskip
6.5mu\scriptstyle+ $}}\subset}}
\newcommand{\sdir}{\ensuremath{\rlap{\raisebox{.15ex}{$\mskip
6.5mu\scriptstyle+ $}}\supset}}
\def\tablecap#1{\vskip 3mm \centerline{#1}\vskip 5mm}
\def\p#1{\partial_#1}
\newcommand{\pd}[2]{\frac{\partial #1}{\partial #2}}
\newcommand{\pdn}[3]{\frac{\partial #1^{#3}}{\partial #2^{#3}}}
\def\DP#1#2{D_{#1}\varphi^{#2}}
\def\dP#1#2{\partial_{#1}\varphi^{#2}}
\def\xh{\hat x}
\newcommand{\Ref}[1]{(\ref{#1})}
\def\ld{\,\ldots\,}

\def\C{{\mathbb C}}
\def\Z{{\mathbb Z}}
\def\R{{\mathbb R}}
\def\mod#1{ \vert #1 \vert }
\def\chapter#1{\hbox{Introduction.}}
\def\Sin{\hbox{sin}}
\def\Cos{\hbox{cos}}
\def\Exp{\hbox{exp}}
\def\Ln{\hbox{ln}}
\def\Tan{\hbox{tan}}
\def\Cot{\hbox{cot}}
\def\Sinh{\hbox{sinh}}
\def\Cosh{\hbox{cosh}}
\def\Tanh{\hbox{tanh}}
\def\Asin{\hbox{asin}}
\def\Acos{\hbox{acos}}
\def\Atan{\hbox{atan}}
\def\Asinh{\hbox{asinh}}
\def\Acosh{\hbox{acosh}}
\def\Atanh{\hbox{atanh}}
\def\frac#1#2{{\textstyle{#1\over #2}}}

\def\ph{\varphi_{m,n}}
\def\phl{\varphi_{m-1,n}}
\def\phr{\varphi_{m+1,n}}
\def\varphil{\varphi_{m-1,n}}
\def\varphir{\varphi_{m+1,n}}
\def\varphit{\varphi_{m,n+1}}
\def\varphib{\varphi_{m,n-1}}
\def\pht{\varphi_{m,n+1}}
\def\phb{\varphi_{m,n-1}}
\def\phbl{\varphi_{m-1,n-1}}
\def\phbr{\varphi_{m+1,n-1}}
\def\phtl{\varphi_{m-1,n+1}}
\def\phtr{\varphi_{m+1,n+1}}
\def\u{u_{m,n}}
\def\ul{u_{m-1,n}}
\def\ur{u_{m+1,n}}
\def\ut{u_{m,n+1}}
\def\ub{u_{m,n-1}}
\def\utr{u_{m+1,n+1}}
\def\ubl{u_{m-1,n-1}}
\def\utl{u_{m-1,n+1}}
\def\ubr{u_{m+1,n-1}}
\def\v{v_{m,n}}
\def\vl{v_{m-1,n}}
\def\vr{v_{m+1,n}}
\def\vt{v_{m,n+1}}
\def\vb{v_{m,n-1}}
\def\vtr{v_{m+1,n+1}}
\def\vbl{v_{m-1,n-1}}
\def\vtl{v_{m-1,n+1}}
\def\vbr{v_{m+1,n-1}}

\def\U{U_{m,n}}
\def\Ul{U_{m-1,n}}
\def\Ur{U_{m+1,n}}
\def\Ut{U_{m,n+1}}
\def\Ub{U_{m,n-1}}
\def\Utr{U_{m+1,n+1}}
\def\Ubl{U_{m-1,n-1}}
\def\Utl{U_{m-1,n+1}}
\def\Ubr{U_{m+1,n-1}}
\def\V{V_{m,n}}
\def\Vl{V_{m-1,n}}
\def\Vr{V_{m+1,n}}
\def\Vt{V_{m,n+1}}
\def\Vb{V_{m,n-1}}
\def\Vtr{V_{m+1,n+1}}
\def\Vbl{V_{m-1,n-1}}
\def\Vtl{V_{m-1,n+1}}
\def\Vbr{V_{m+1,n-1}}
%%%%
\def\tr{{\rm tr}\,}
% Greek Letter

\def\a{\alpha}
\def\b{\beta}
\def\g{\gamma}
\def\d{\delta}
\def\ep{\epsilon}
\def\e{\varepsilon}
\def\z{\zeta}
\def\t{\theta}
\def\k{\kappa}
\def\l{\lambda}
\def\s{\sigma}
\def\f{\varphi}
\def\w{\omega}
\def\v{{\hbox{v}}}
\def\u{{\hbox{u}}}
\def\x{{\hbox{x}}}
%%%%%%%%%%%%%%%%%%%%%%%%

\newcommand{\ie}{{\it i.e.}}
\newcommand{\cmod}[1]{ \vert #1 \vert ^2 }
\newcommand{\cmodn}[2]{ \vert #1 \vert ^{#2} }
\newcommand{\nhat}{\mbox{\boldmath$\hat n$}}
\nopagebreak[3]
\bigskip

\title{ \bf Veronese subsequent analytic solutions of the $\mathbb{C}P^{2s}$ sigma model equations described via Krawtchouk polynomials}
\vskip 1cm

\bigskip
\author{
Nicolas Cramp\'e\thanks{email address: crampe1977@gmail.com}\\
Institut Denis Poisson, Universit\'e de Tours -- Universit\'e d'Orl\'eans\\
Parc de Grandmont, 37200 Tours, France \acc 
Alfred Michel Grundland\thanks{email address: grundlan@crm.umontreal.ca}
\\
Centre de Recherches Math{\'e}matiques, Universit{\'e} de Montr{\'e}al,\\
C. P. 6128, Succ.\ Centre-ville, Montr{\'e}al, (QC) H3C 3J7,
Canada\\ Universit\'{e} du Qu\'{e}bec, Trois-Rivi\`{e}res, CP500 (QC) G9A 5H7, Canada} \date{}

\maketitle
\vspace{-5mm}
\begin{abstract}\vspace{-3mm}
The objective of this paper is to establish a new relationship between the Veronese subsequent analytic solutions of the Euclidean $\mathbb{C}P^{2s}$ sigma model in two dimensions and the orthogonal Krawtchouk polynomials. We show that such solutions of the $\mathbb{C}P^{2s}$ model, defined on the Riemann sphere and having a finite action, can be explicitly parametrised in terms of these polynomials. We apply the obtained results to the analysis of surfaces associated with $\mathbb{C}P^{2s}$ sigma models, defined using the generalized Weierstrass formula for immersion. We show that these surfaces are spheres immersed in the $\mathfrak{su}(2s+1)$ Lie algebra, and express several other geometrical characteristics in terms of the Krawtchouk polynomials. Finally, a new connection between the $\mathfrak{su}(2)$ spin-s representation and the $\mathbb{C}P^{2s}$ model is explored in detail. It is shown that for any given holomorphic vector function in $\mathbb{C}^{2s+1}$ written as a Veronese sequence, it is possible to derive subsequent solutions of the $\mathbb{C}P^{2s}$ model through algebraic recurrence relations which turn out to be simpler than the analytic relations known in the literature.
\end{abstract}
\vspace{-4mm}
%Short Title: 
\begin{center}{In honour of Decio Levi (University of Roma Tre)}\end{center}
%PACS: 03.40.Kf, 02.20.Sv, 02.30.Jr
\vspace{-3mm}


\vspace{3mm}

\newpage

\section{The $\mathbb{C}P^{2s}$ sigma model}
 
The dynamical fields in the $\mathbb{C}P^{2s}$ sigma models are maps from the Riemann sphere $\mathbb{S}^{2}$ to the complex projective space $\mathbb{C}P^{2s}\simeq \mathbb{S}^{4s(s+1)}/U(1)$
\begin{displaymath}
\mathbb{S}^2\ni\xi_\pm=\xi^1\pm i \xi^2 \mapsto z=(z_0,z_1,\dots,z_{2s})\in\mathbb{C}^{2s+1}\setminus\{\emptyset\},
\end{displaymath}
which are stationary points of the action functional \cite{Zakrzewski}
\begin{equation}\label{eq1}
\mathcal{A}=\frac{1}{4}\iint_{\mathbb{S}^2} (D_{\mu}z)^{\dagger}\cdot(D_{\mu}z) d\xi_+ d\xi_-,
\end{equation}
and hence are solutions of the Euler-Lagrange (EL) equations
\begin{equation}\label{eq2}
D_{\mu}D_{\mu}z+(D_{\mu}z)^{\dagger}\cdot(D_{\mu}z)z=0,
\end{equation}
subjected to $z^\dagger z=1$, where $D_{\mu}$ are the covariant derivatives defined by
\begin{displaymath}
D_{\mu}z=\partial_{\mu}z-(z^{\dagger}\partial_{\mu}z)z,\qquad \partial_{\mu}=\frac{\partial}{\partial\xi^{\mu}},\qquad \mu=1,2.
\end{displaymath}
We require that the action (\ref{eq1}) over the whole Riemann sphere $\mathbb{S}^{2}$ be finite.

\section{Projective formalism}

Equivalently, representing the z's by their homogeneous representatives, i.e. maps into $\mathbb{C}^{2s+1}\setminus\{\emptyset\}$
\begin{displaymath}
z=\frac{f}{(f^{\dagger}\cdot f)^{1/2}}\cdot
\end{displaymath}
We may use (fields of) rank-1 Hermitian projectors
\begin{equation}\label{eq3}
P=\frac{f\otimes f^\dagger}{f^\dagger\cdot f}, \qquad P^2=P,\qquad P^\dagger=P.
\end{equation}
This places the EL equations into the form of the conservation law (CL)
\begin{equation}\label{eq4}
\partial[\bar{\partial}P,P]+\bar{\partial}[\partial P,P]=0,
\end{equation}
where the symbols $\partial$ and $\bar{\partial}$ stand for the complex derivatives with respect to $\xi_+$ and $\xi_-$ given by
\begin{displaymath}
\partial=\frac{1}{2}\left(\frac{\partial}{\partial\xi^1}-i\frac{\partial}{\partial\xi^2}\right),\qquad \bar{\partial}=\frac{1}{2}\left(\frac{\partial}{\partial\xi^1}+i\frac{\partial}{\partial\xi^2}\right).
\end{displaymath}
Under the above assumptions every solution can be obtained from a holomorphic (respectively antiholomorphic) solution
\begin{displaymath}
f:\mathbb{S}^2\rightarrow\mathbb{C}^{2s+1}\setminus\{\emptyset\},\qquad \bar{\partial}f=0,
\end{displaymath}
by successive applications of the raising or lowering operator \cite{Zakrzewski},
\begin{equation}\label{eq5}
f_{k+1}=P_+(f_k):=(\mathbb{I}_{2s+1}-P_k)\partial f_k,\qquad f_{k-1}=P_-(f_k):=(\mathbb{I}_{2s+1}-P_k)\bar{\partial}f_k,
\end{equation}
\begin{displaymath}
P_{\pm}^0=\mathbb{I}_{2s+1},\qquad P_{\pm}^{2s+1}f_k=0,\qquad k=0,1,\ldots,2s,
\end{displaymath}
where $P_+(f_k)$ is a creation operator and $P_-(f_k)$ is an annihilation operator. Thus the sequence of solutions in the $\mathbb{C}P^{2s}$ model consists of $2s+1$ vectors $f_k$ or  $2s+1$ rank-1 Hermitian projectors $P_k$. The action integral (\ref{eq1})  in terms of the 
projectors $P_k$ has the form
\begin{equation}\label{eq6}
\mathcal{A}(P_k) = \iint_{\mathbb{S}^2}tr\left(\partial P_k\cdot\bar{\partial} P_k\right)d\xi_+ d\xi_-.
\end{equation}
In terms of the nonconstant projectors $P_k$, the recurrence relations (\ref{eq5}) become \cite{Goldstein1}
\begin{equation}\label{eq7}
P_{k\pm 1} = \Pi_\pm(P_k) :=  \frac{(\partial_{\pm} P_k)P_k(\partial_{\mp} P_k)}{tr[(\partial_{\pm} P_k)P_k(\partial_{\mp} P_k)]},
\end{equation}
for $\;\; tr[(\partial_{\pm} P_k)P_k(\partial_{\mp} P_k)]\neq 0 \;$ and are equal to zero when $tr[(\partial_{\pm} P_k)P_k(\partial_{\mp} P_k)]=0$, where $\partial_+$ and $\partial_-$ stand for $\partial$ and $\bar{\partial}$, respectively. Here $P_k$ stands for one of the projectors $\{P_0,P_1,\ldots,P_{2s}\}$. This set satisfies the orthogonality and completeness relations
\begin{equation}\label{eq8}
P_jP_k = \delta_{jk}P_j,\quad 0\leq k,j\leq 2s,\qquad \sum_{j = 0}^{2s}P_j = \mathbb{I}_{2s+1}.
\end{equation}\hspace{1mm}

\section{Solutions of the $\mathbb{C}P^{2s}$ sigma model}
A particular holomorphic solution of the $\mathbb{C}P^{2s}$ model equations (\ref{eq4}) expressed in terms of the f's
\small\begin{equation}\label{eq9}
\left(\mathbb{I}_{2s+1}-\frac{f_k\otimes f_k^\dagger}{f_k^\dagger\cdot f_k}\right)\left[\partial\bar{\partial}f_k-\frac{1}{f_k^\dagger\cdot f_k}\left(\left(f_k^\dagger\cdot\bar{\partial}f_k\right)\partial f_k+\left(f_k^\dagger\cdot\partial f_k\right)\bar{\partial}f_k\right)\right]=0,\end{equation} \normalsize
for $0\leq k\leq 2s$, can be written as the Veronese sequence \cite{Bolton}
\begin{equation}\label{eq10}
f_0=\left(1,\binom{2s}{1}^{1/2}\xi_+,\dots,\binom{2s}{r}^{1/2}\xi_+^r,\dots,\xi_+^{2s}\right)\in\mathbb{C}^{2s+1}\backslash\{\emptyset\},\qquad \mbox{for }k=0.
\end{equation}
Subsequent solutions of (\ref{eq9}) can be obtained by acting with the creation operators (\ref{eq5}). Thus for $k>2$ this procedure allows us to construct three classes of solutions: holomorphic $f_0$,
antiholomorphic $f_{2s}$ and mixed solutions $f_k$, $1\leq k\leq 2s-1$.\\\\
Under the above assumptions we show that any rank-1 Hermitian projector solution $P_k$ of the EL equations (\ref{eq9}) can be expressed explicitly in terms of the Krawtchouk orthogonal polynomials \cite{Crampe}.

\noindent{\bf Theorem 1 (The main result).}
Let the $\mathbb{C}P^{2s}$ model be defined on the Riemann sphere  $\mathbb{S}^{2}$ and have a finite action functional. Then the Veronese subsequent analytic solutions $f_k$ of the $\mathbb{C}P^{2s}$
model (\ref{eq9}) take the form
\begin{equation}\label{eq11}
(f_k)_{j} = \frac{(2s)!}{(2s-k)!}\left(\frac{-\xi_-}{ 1+\xi_+\xi_-}\right)^k \sqrt{\binom{2s}{j}}\xi_+^jK_j(k; p, 2s), \qquad 0\leq k,j\leq 2s
\end{equation}
\begin{displaymath}
0<p = \frac{\xi_+\xi_-}{1+\xi_+\xi_-}<1,
\end{displaymath}
where $(f_k)_j$is the jth component of the vector $f_k \in\mathbb{C}^{2s+1}\setminus\{\emptyset\}$ and $K_j(k;p,2s)$ are Krawtchouk orthogonal polynomials defined in terms of the hypergeometric function
\begin{equation}\label{eq12}
K_j(k)=K_j(k; p, 2s) = %\prescript{}
_{2}\hspace{-1mm}F_{1}(-j, -k; -2s; 1/p),\quad 0\leq k\leq 2s.
\end{equation}
Here $j$,$k$ and $2s$ are parameters, $p$ is an argument in (\ref{eq12}). We use the convention
\begin{equation}\label{eq13}
K_j(0;p,2s)=1,\quad \mbox{for } k=0.
\end{equation}
The vectors $f_k$ can be used to construct the rank-1 Hermitian matrix projector $P_k$ with an entry in the $i^{th}$ row and $j^{th}$ column given by
\begin{equation}\label{eq14}
(P_k)_{ij} = \binom{2s}{k}\frac{(\xi_+\xi_-)^{k}}{(1+\xi_+\xi_-)^{2s}}\xi_+^i\xi_-^j\sqrt{\binom{2s}{i}\binom{2s}{j}}K_i(k)K_j(k),
\end{equation}
where, in what follows, we use the following abbreviated notation
\begin{equation}\label{eq15}
K_j(k) :=K_j(k; p, 2s), \qquad  K_j(k\pm1):=K_j(k\pm1; p, 2s).
\end{equation}
The EL equations (\ref{eq9}) with the idempotency condition $P_k^2=P_k$ admit a larger class of solutions than the rank-1 Hermitian projector $P_k$ \cite{Goldstein4}.

\noindent{\bf Proposition 1 (Higher-rank projectors).}
Let the linear combinations of rank-1 Hermitian projectors $P_l$ be
\begin{equation}
P = \sum_{l=0}^{2s}\lambda_l P_l, \qquad \lambda_l = 0\;\;\text{ or }\;\; \lambda_l = 1 \;\;\text{ for \; all }\;\;l\in\{0, 1, ..., 2s\},
\end{equation}
for which $P_l$ satisfy the the EL equations (\ref{eq9}). The higher-rank projector $P$ can be expressed in terms of the Krawtchouk polynomials
\begin{equation}\label{eq16}
(P)_{ij} = \sum_{l=0}^{2s}\lambda_l\binom{2s}{l}\frac{(\xi_+\xi_-)^l}{(1+\xi_+\xi_-)^{2s}}\xi_-^i\xi_+^j\sqrt{\binom{2s}{i}\binom{2s}{j}}K_i(l)K_j(l)
\end{equation}
which satisfy both the EL equations (\ref{eq9}) and the idempotency condition $P^2=P$. In this case the projector $P$ maps the $\mathbb{C}^{2s+1}$ space onto $\mathbb{C}^k$, where $k=\sum_{l=0}^{2s}\lambda_l$.

\section{The  $\mathfrak{su(2)}$ spin-s representation} 
A direct connection was established between the  $\mathbb{C}P^{2s}$ model and the spin-s  $\mathfrak{su(2)}$ representation \cite{Goldstein2,Goldstein3}. The spin matrix $S^z$ is defined as a linear combination of the
(2s+1) rank-1 Hermitian projectors $P_k$, i.e.
\begin{equation}\label{eq17}
S^z(\xi_+,\xi_-)=\sum_{k=0}^{2s}(k-s)P_k,\qquad (S^z)^{\dagger}=S^z,
\end{equation}
where the eigenvalues of the generator $S^z$ are $\{-s,-s+1,\dots,s-1,s\}$. They are either integer (for odd $2s+1$) or half-integer (for even $2s+1$) values. From equation (\ref{eq17}) we obtain that the spin matrix $S^z$ is given by the tridiagonal matrix with an entry in the $i^{th}$ row and $j^{th}$ column \cite{Crampe}
\begin{equation}\nonumber
(S^z)_{ij} = \delta_{ij}\left(\frac{1-\xi_+\xi_-}{1+\xi_+\xi_-}\right)(i-s) - \delta_{i-1,j}\left(\frac{\xi_+}{1+\xi_+\xi_-}\right)\sqrt{i(2s+1-i)}
\end{equation}
\begin{equation}\label{eq18} 
- \delta_{i,j-1}\left(\frac{\xi_-}{1+\xi_+\xi_-}\right)\sqrt{j(2s-j+1)},\qquad 0\leq i, j \leq 2s.
\end{equation}
The generators $S^z$ and  $S^{\pm}$ of the  $\mathfrak{su}(2)$ Lie algebra satisfy the commutation relations
\begin{equation}\label{eq19}
[S^z, S^\pm]=\pm S^\pm,\qquad [S^+, S^-] = 2S^z,
\end{equation}
and they are identified with the following $(2s+1)\times(2s+1)$ matrices \cite{Merzbacher}\\~\\
\begin{align}\label{eq20}
(\sigma^z)_{ij} &= (s-i)\delta_{ij},\;\\
(\sigma^+)_{ij} &= \sqrt{(2s-j+1)j}\delta_{i,j-1}, \qquad\qquad 0\leq i, j \leq 2s\\
(\sigma^-)_{ij} &=  \sqrt{(2s-i+1)i}\delta_{i-1,j}.
\end{align}\\~\\
Hence the matrices $S^z$ and  $S^{\pm}$ can be decomposed as a linear combination of the matrices $\sigma^z$ and $\sigma^{\pm}$, namely
\begin{equation}\label{eq21}
\begin{pmatrix} S^z \\ S^+ \\ S^- \end{pmatrix}=\frac{1}{1+\xi_+\xi_-}\begin{pmatrix} \xi_+\xi_- -1 & -\xi_- & -\xi_+ \\ 2\xi_- & \xi_-^2 & -1 \\ 2\xi_+ & -1 & \xi_+^2\end{pmatrix}\begin{pmatrix} \sigma^z \\ \sigma^+ \\ \sigma^- \end{pmatrix},
\end{equation}
where $(S^+)^{\dagger}=S^-$ and $(S^-)^{\dagger}=S^+$. The eigenvalue problem for the spin matrix $S^z$ is given by
\begin{displaymath}
S^zf_k=(k-s)f_k,\qquad S^z(S^{\pm}f_k)=(k\pm 1-s)(S^{\pm}f_k),\qquad\mbox{for}\quad 0\leq k\leq 2s.
\end{displaymath}
Under these circumstances the following holds

\noindent{\bf Proposition 2 (Recurrence relations associated with the $\mathbb{C}P^{2s}$ models).}
For the subsequent Veronese analytic solutions $f_k$ of the $\mathbb{C}P^{2s}$ model (\ref{eq9}), the algebraic recurrence relations for the vectors $S^zf_k$ and $S^{\pm}f_k$ are given by
\begin{equation}\label{eq22}
S^+f_k =\begin{cases} -(1+\xi_+\xi_-)f_{k+1}\quad \text{for}\quad 0 \leq k\leq 2s-1, \\
 0 \quad \text{for}\quad k=2s,
\end{cases}
\end{equation}
\begin{equation}\label{eq23}
S^-f_k = \frac{1}{1+\xi_+\xi_-}k(k-1-2s)f_{k-1} \quad \mbox{for}\quad 0 \leq k\leq 2s.
\end{equation}
In terms of the projectors $P_k$, these recurrence relations (\ref{eq7}) take the form
\begin{equation}\label{eq24}
P_{k+1}=\Pi_+(P_k):=\frac{S^+P_kS^-}{tr(S^+P_kS^-)},\qquad P_{k-1}=\Pi_-(P_k):=\frac{S^-P_kS^+}{tr(S^-P_kS^+)},
\end{equation}
where $tr(S^+P_kS^-)\neq 0$.

The proof of the formulae (\ref{eq24}) follows immediately from (\ref{eq3}) since the relations (\ref{eq22}) and (\ref{eq23}) hold. Note that the relations (\ref{eq22}) and (\ref{eq23}) allow us to recursively construct the subsequent solutions $f_k$  from the holomorphic solution $f_0$ in a simpler way than the ones obtained from the analytic recurrence relation (\ref{eq5}). Therefore, the matrices $S^{\pm}$ are the creation and annihilation operators for the vectors $f_k$ and the projectors $P_k$. The result given in the above proposition can be interpreted as the matrix elements of the $SU(2)$ irreductible representations, known as the Wigner D function. It is known \cite{Koornwinder,Granovski} that these matrix elements can be expressed in terms of the Krawtchouk polynomials.

\section{Geometrical aspects of surfaces}

The generalised Weierstrass formula for the immersion of 2D-surfaces associated with the $\mathbb{C}P^{2s}$ model (\ref{eq9}) is given by \cite{Grundland}
\begin{equation}\label{eq25}
X_k(\xi_+,\xi_-)=-i\left(P_k+2\sum_{j=0}^{k-1}P_j\right)+i\left(\frac{1+2k}{1+2s}\right)\mathbb{I}_{2s+1} \in \mathfrak{su}(2s+1).
\end{equation}
For the sake of uniformity, the inner product is defined by
\begin{displaymath}
(A,B)=-\frac{1}{2}tr(A\cdot B),\qquad A,B\in \mathfrak{su}(2s+1).
\end{displaymath}
The first and second fundamental forms are\\~\\
\begin{align}\label{eq26}
I_k &= tr(\partial P_k\cdot \bar{\partial} P_k)d\xi_+d\xi_- = \frac{2(2sk+s-k^2)}{(1+\xi_+\xi_-)^2}d\xi_+d\xi_-,\nonumber\\
 II_k  &= -tr(\partial P_k\cdot \bar{\partial} P_k) \partial \left(\frac{[\partial P_k,P_k]}{tr(\partial P_k\cdot \bar{\partial} P_k)} \right) d\xi_+^2 +2i [\bar{\partial} P_k, \partial P_k]d\xi_+d\xi_-\\
 &\qquad\qquad-tr(\partial P_k\cdot \bar{\partial} P_k) \bar{\partial} \left( \frac{[\bar{\partial} P_k,P_k]}{tr(\partial P_k\cdot \bar{\partial} P_k)} \right) d\xi_-^2.\nonumber
\end{align}

\noindent{\bf Proposition 3 (non-intersecting spheres).}
For any value of the Veronese subsequent analytic solutions $f_k$ of the $\mathbb{C}P^{2s}$ model (\ref{eq9}), all the 2D-surfaces $X_k$ are non-intersecting spheres with the radius
\begin{equation}\label{eq27}
R_k=(X_k, X_k)^{1/2} = \left(-\frac{1}{2}tr(X_k)^2\right)^{1/2} = \Bigg{|}\frac{-2k^2+2k(2s-1)+s-1}{1+2s}\Bigg{|}^{1/2},
\end{equation}
immersed in the Lie algebra $\mathfrak{su}(2s+1)\simeq \mathbb{R}^{4s(s+1)}$.

Outline of the proof: Let us assume that $l>k$ are two different indices of the induced surfaces. Substracting (\ref{eq25}) from the analogous expression for $X_l$, we get
\begin{equation}\label{eq28}
P_l-P_k+2\sum_{j=k}^{l-1}P_j-\frac{2(l-k)}{2s+1}\mathbb{I}_{2s+1}=0.
\end{equation}
Multiplying equation (\ref{eq28}) by $P_k$, $P_l$ or $P_{l-1}$ and solving the obtained system of equations, we obtain that the 2D-surfaces $X_k$ and $X_l$ do not intersect if $k\neq l$ with the exceptions of $X_0$ and $X_1$
in the  $\mathbb{C}P^{1}$ model since $X_0$ and $X_1$ coincide. The fundamental forms (\ref{eq26}) imply that the Gaussian curvatures of the 2D-surfaces have constant positive values
\begin{equation}\label{29}
K_k=\frac{2}{2sk+s-k^2}.
\end{equation}
The K\"ahler angles are given by
\begin{displaymath}
\tan{\left(\dfrac{1}{2}\theta_k(m)\right)}=\Bigg{|}\dfrac{df_k(m)(\partial/\partial\xi_-)}{df_k(m)(\partial/\partial\xi_+)}\Bigg{|},\qquad m\in\mathbb{S}^2
\end{displaymath}
and have constant positive values
\begin{displaymath}
\cos{\theta_k}=\dfrac{s-k}{2sk+s-k^2}.
\end{displaymath}
The Euler-Poincar\'e characters of the 2D-surfaces $X_k$ are the integer $\Delta_k=2$ for all k such that $0\leq k\leq 2s$. This means that all 2D-surfaces associated with the  $\mathbb{C}P^{2s}$ model are
non-intersecting spheres with radius $R_k$ given by (\ref{eq27}).$\left.\right.$\hfill$\square$

\noindent The technique for obtaining surfaces via projective structures and their links with orthogonal polynomials, elaborated from the  $\mathbb{C}P^{2s}$ models, can be extended to different types of Grassmannian manifolds. An analysis of these manifolds can provide us with much more diverse types of surfaces.

\noindent {\bf Acknowledgements}\\
This research was supported by the NSERC operating grant of one of the authors (A.M.G.). N.C. is indebted to the Centre de Recherches Math\'ematiques (CRM), Universit\'e de Montr\'eal for the opportunity to hold a CRM-Simons professorship.

{}

\end{document}